\documentclass[10pt,preprint,twocolumn]{article}
\usepackage{authblk}
\usepackage[utf8]{inputenc}
\usepackage{amssymb}
\usepackage{amsmath,amsthm,pifont,amsfonts,MnSymbol}
\usepackage{graphicx}
\usepackage{hyperref}
\usepackage[a4paper, portrait, margin=1in]{geometry}

\begin{document}

\title{Chronofold: a data structure for versioned text}

\author{Victor Grishchenko}

\author{Mikhail Patrakeev}

\affil{
	N.N. Krasovskii Institute of Mathematics and Mechanics,
	Ekaterinburg,
	Russia
}

\maketitle

\begin{abstract}
Chronofold is a replicated data structure for versioned text.
It is designed for use in collaborative editors and revision control systems.
Past models of this kind either retrofitted local linear orders to a distributed system (the OT approach) or employed distributed data models locally (the CRDT approach).
That caused either extreme fragility in a distributed setting or egregious overheads in local use.
Overall, that local/distributed impedance mismatch is cognitively taxing and causes lots of complexity.
We solve that by using subjective linear orders locally at each replica, while inter-replica communication uses a distributed model.
A separate translation layer insulates local data structures from the distributed environment.
We modify the Lamport timestamping scheme to make that translation as trivial as possible.
We believe our approach has applications beyond the domain of collaborative editing.

\end{abstract}

\theoremstyle{plain}
\newtheorem{teor}{Theorem}[section]
\newtheorem{lemm}[teor]{Lemma}
\newtheorem{corr}[teor]{Corollary}
\newtheorem{prop}[teor]{Proposition}
\newtheorem{rema}[teor]{Remark}
\theoremstyle{definition}
\newtheorem{defi}[teor]{Definition}
\newtheorem{term}[teor]{Terminology}
\newtheorem{nota}[teor]{Notation}
\newtheorem{conv}[teor]{Convention}
\theoremstyle{remark}
\newtheorem{exam}[teor]{Example}
\newtheorem{ques}[teor]{Question}

\mathsurround=1pt
\newcommand{\nos}{\mathsurround=0pt}
\newcommand{\NN}{\mathbb{N}}
\newcommand{\PROC}{\mathsf{PROC}}
\newcommand{\proc}{\mathsf{proc}}
\newcommand{\auth}{\mathsf{auth}}
\newcommand{\andx}{\mathsf{andx}}
\newcommand{\w}{\mathsf{w}}
\newcommand{\ndx}{\mathsf{ndx}}
\newcommand{\sh}{\mathsf{shft}}
\newcommand{\vl}{\mathsf{val}}
\newcommand{\rf}{\mathsf{ref}}
\renewcommand{\lg}{\mathsf{log}}
\newcommand{\LG}{\mathsf{LOG}}
\renewcommand{\ll}{\langle}
\newcommand{\rr}{\rangle}
\renewcommand{\iff}{\quad{\colon}{\longleftrightarrow}\quad}
\newcommand{\pred}{\mathsf{pred}}
\renewcommand{\succ}{\mathsf{succ}}
\newcommand{\lh}{\mathsf{lh}}
\newcommand{\op}{\mathsf{op}}
\overfullrule0pt

\section{ Introduction }

Even without the real-time collaboration, data structures for editable text is a vast field on its own.
Plain text storage and transmission is not a challenge for modern computers;
``War and Peace'' weighs 3MB, on par with a smartphone photograph.
Text editing is more demanding, as it needs fast writes and some basic versioning functionality (at least, to support undo/redo).
Naive implementations do not suffice;
there is an entire class of editable-text data structures, such as gap buffers~\cite{seq}, piece tables~\cite{piece}, ropes~\cite{ropes} and others.
Integrated Development Environments (IDEs) have even more reasons to version the edited text;
one of them is asynchronous communication between multiple worker threads or processes.
Finally, there are Source Code Management systems (SCM), where texts are versioned and stored long-term.
The underlying models of text versioning have plenty of overlap in these three classes of applications.

The classic plain text versioning model sees any document change (a \emph{diff}) as a number of range insertions and deletions.
Alternatively, that can be generalized to a number of range replacements (\emph{splices}).
The Myers algorithm~\cite{myers1986ano} can calculate a diff from two versions of a text in $O(ND)\nos$ time,
where $N$ is the combined size of the texts and $D$ is the size of the changes.
Thus, the worst case is $O(N^2).$
That is less of a problem for \texttt{diff}, \texttt{patch}, \texttt{svn}, \texttt{git}, etc, as
their unit of change is a line of text.
There are much less lines than characters and lines are more unique, so a number of optimizations and heuristics make Myers good enough in all the reasonable cases.
If the unit of change is a \emph{character}, Myers is much more of a challenge;
e.g. Google's diff-match-patch~\cite{diff-match-patch} library uses timers to provide a good-enough result in acceptable time.
Another issue with the diff approach is its non-determinism in case of concurrent changes.
To integrate a change, \texttt{patch} relies on its position and \emph{context} (the text around the changed spot).
Concurrent changes may garble both, causing mis-application of a patch.
Partially, that is solved by heuristics. Still, SCMs require manual merge of changes in any non-trivial cases.

\emph{Weave}~\cite{rochkind1975source} is a classic data structure for text versioning.
It was invented in the 1970s and used in SCCS, TeamWare and most recently BitKeeper as the main form of storage
and by many other SCMs for merge of concurrent changes.
A weave has a reputation of one of the most reinvented data structures in history.
It is alternatively known as "interleaved deltas", "union string", and under other names.
Its key idea is simple: annotate all pieces of a text (\emph{deltas}) with their ``birth'' and ``death'' dates,
keep the deleted pieces in their place.
Then, one pass of such a collection can produce any version of the text, if all the ``dead'' and ``yet-unborn'' pieces are filtered out.
The top issue with a weave is that it needs to be spliced on every edit (i.e. rewritten in full), very much like a plain string.	
The original SCCS weave was line-based, but we will use that as a broad term for this kind of a data structure, no matter line- or character-based.

Notably, the widely popular \texttt{git} SCM~\cite{git} has immutable binary blobs as its primary abstraction, no deltas.
Still, it employs delta-based data structures to merge concurrent changes, while its internal storage format is organized around delta compression.
It also supports line-based patches and blame maps.
Ironically, declaring blobs as its primary abstraction made \texttt{git} use deltas more, not less.

The \emph{Operational Transformation} (OT) model~\cite{Ellis89concurrencycontrol} originated from the first experiments with real-time collaborative editors in the 80s.
With OT, each single-character edit can be sent out immediately as an \emph{operation} (op).
OT needed deterministic merge of changes, despite any concurrent modification.
Hence, it relied on positions, not contexts, to apply the changes.
Positions are also affected by concurrent edits, so OT iteratively \emph{transforms} the operations  to keep them correct.
That works reasonably well, except that concurrent modifications create combinatorially complex and highly counter-intuitive effects.
For that reason, any practical OT implementation relies on a central server to transform the ops.
Despite its somewhat torturous history, OT eventually led to such applications as Google Docs.
	
In 2006, the dissatisfaction with OT led to a new proposal named WithOut Operational Transforms (WOOT)~\cite{woot}.
Its cornerstone change was to assign every character a unique identifier (id).
Then, WOOT represents a text as a directed acyclic graph of characters, each one referencing its left and right neighbors at the time of insertion.
The order of identifiers resolves ties between concurrent insertions to the same location.
Deleted characters get marked with \emph{tombstones}.
WOOT ops are immutable and commutative, hence immune to reordering issues.
While obviously correct, WOOT was highly impractical due to metadata overheads.

\emph{Causal Tree} (CT)~\cite{grishchenko_2010} aimed at improving WOOT, along with Logoot~\cite{weiss_urso_molli_2010}, TreeDoc~\cite{Preguica}, LSEQ~\cite{LSEQ} and other proposals.
In particular, CT reduces per-character metadata to (a) \emph{logical timestamp} (b) timestamp of the preceding character.
Logical timestamps~\cite{lamport2019time} are tuples $\ll t, \alpha \rr$ where $t$ is the logical time value and $\alpha$ is the process id.
The lexicographic order of timestamps forms the \emph{arbitrary total order}~\cite{lamport1978ordering} (ATO) consistent with the cause-effect ordering (``happened-before'').
CT employed fixed-width logical timestamps of various kinds, while Logoot and TreeDoc used variable-length identifiers.
Once the first CT draft~\cite{grishchenkocausal} appeared in 2008, it was immediately noted~\cite{bramc} that CT's inner workings are very reminiscent of a weave.
In 2010, the Replicated Growable Array (RGA)~\cite{ROH2011354} algorithm was proposed.
In 2016, it was formally proven~\cite{attiya2016specification} that RGA and CT use the same algorithm
(curiously, the paper uses another term for CT, a Timestamped Insertion Tree).
Interestingly, OT-with-tombstones proposals~\cite{TTF,raphot} resulted in similar weave-based algorithms.

In 2009, the authors of TreeDoc proposed a broad term for this kind of commutativity/convergence based algorithms: Conflict-free Replicated Data Types (CRDT)~\cite{mihai2009crdts,shapiro2011conflict}.
Although the potential of commutative ops was noticed as far back as 1987~\cite{Schneider}, only now the topic had close attention.
But, despite both industry and academia making circles around CRDTs, no standard solution emerged yet.
The key issue remained the same: metadata overheads and cognitive costs of a distributed data model.
So far, industry adoption of CRDTs had the air of a pilot project.
By 2013, CT was deployed in the Yandex Live Letters collaborative editor~\cite{grishchenko2014citrea, papyrus} which was phased out several years later.
Another CRDT based editor was made in Yandex~\cite{ya2} by 2019.
The Atom editor employed CRDT~\cite{tt} for its collaborative features.
Apple Notes is known to sync with CRDTs~\cite{applenotes}.
Some used simplified ersatz-CRDT models~\cite{ersatz}.
As recently as in 2019, two high-profile CRDT-based editor projects fell short of the objectives (Google-associated \texttt{xi}~\cite{xi} and GitHub's \texttt{xray}~\cite{xray}).
Authors cited data structure complexity as a major impediment.

CRDT overheads and complexity are rooted in the use of per-character ids and the need to map them to positions in the weave and the text.
That data has to be stored in addition to the regular editable-text data structures.
According to the original article, RGA needs a hash map to store a linked list, one entry per character.
A naive CT implementation may store a text as an actual tree of letters, with each letter being a rather complex object.
Such implementations are known to exist and they don't work well.
Overall, turning a character into an object brings lots of overheads: pointers, headers, cache misses, garbage collection, etc.
Every implementation tried to optimize that, one way or another.

Some tried to work with blocks of characters~\cite{andre2013supporting,briot2016high} instead of single characters;
e.g. add a copy-pasted fragment as a single op, then split it later if necessary.
Another approach was to compress ranges of timestamps~\cite{ron1} which are close in value.
The \texttt{xi} editor used a hybrid OT-CRDT approach (``coordinate transforms'') to save on the metadata;
predictably, that increased algorithm and data structure complexity.
CT survived multiple major revisions addressing the issue of overheads.
Most CT implementations used flat data structures: strings, arrays and buffers~\cite{laced,ron1}.
In particular, the 2012 JavaScript implementation~\cite{grishchenko_2010,grishchenko2014citrea} used a peculiar optimization technique coding timestamps as tuples of characters and using regex scans to avoid keeping a per-symbol hash map.
The 2017 RON CT implementation~\cite{ron1} borrowed the iterator heap technique~\cite{cassandra} from LSMT databases.
It merges $i$ inputs by a single $O(N \log i)$ pass; the inputs might be versions, patches and/or single ops.
The technique is perfect for batched server-side operations, not so much for real-time client-side use.
So far, optimizations did not lower CRDT overheads to the level of a piece table or at least comparable.
That makes CRDTs acceptable for niche uses, but not as a general-purpose data structure for versioned text.

This paper proceeds as follows.
Sec.~\ref{logtime} explains the category of subjective linear orders and log timestamps, a logical timestamping scheme.
That is the key to the article as it lets us use linear addressing in a distributed data structure.
In Sec.~\ref{chronofold}, we introduce chronofold, a data structure for versioned text.
Further, in Sec.~\ref{costruct} we put chronofold into the wider context of a complex editor or revision control system and explain how it works in lockstep with other data structures.
Finally, Sec.~\ref{done} concludes with our findings.

\section{CT, log time and subjective orders} \label{logtime}

In distributed systems, events happen ``fast'' while messages propagate ``slowly''.
As a result, the perceived order of events is different for different observers.
No wonder the seminal paper on distributed systems~\cite{lamport1978ordering} drew inspiration from relativistic physics;
its key concept of \emph{logical time} is dependent on the frame of reference.

The CT model is defined in a way to be independent of any replica's local perceived operation order (\emph{subjective} order).
CT works in terms of a \emph{causal} partial order of operations and a compatible timestamp-based ATO.
That makes the model simple and its behavior self-evident.
The unfortunate cost is that addressing, data structures, and versioning become non-linear and thus complex.

We found that the inner workings of the system might be greatly simplified if we rely on those linear subjective orders instead of ignoring them.
As long as the system produces the same text, we have the best of both worlds: simplicity of linear addressing and resilience of a distributed model.
In the Replicated Causal Tree model (RCT) we make subjective orders explicit and consider their properties.

We denote \emph{processes} by variables $\alpha\nos,$ $\beta\nos,$ $\gamma\nos;$ the variables ${i}\nos,$ ${j}\nos,$ ${k}\nos,$ ${m}\nos,$ ${n}\nos$ range over the set of natural numbers $\{1,2,\ldots\}.$
Processes create and exchange \emph{operations} (ops) identified by \emph{timestamps} (ids).
A timestamp is an ordered pair $t = \ll{n},\alpha\rr$
where $\alpha$ is the \emph{author} also denoted by $\auth({t}),$ and ${n}$ is the \emph{author's index} also denoted by $\andx({t}).$
An \emph{op} is a tuple $\ll t, \rf(t), \vl(t)\rr$ where $t$ is its id, $\rf({t})$ is the id of its CT parent, and $\vl({t})$ is its value (a character).
Each process $\alpha$ keeps a subjectively ordered \emph{log} of ops it either authored 
 or received from other processes.

\begin{defi} \label{main.def}
  \emph{Replicated  Causal Tree, RCT,} is a tuple ${R}=\ll{T},\vl,\rf,\lg\rr$ such that ${T}$ is a set of timestamps, $\vl$ is a function with domain ${T},$
  $\rf$ is a function from ${T}$ to ${T},$ and $\lg$ is a function from the set
  $\proc({R})\coloneq\{\auth({t}):{t}\in{T}\}$
  to the set of injective sequences in ${T},$ which associates to every process $\alpha\in\proc({R})$ the sequence $\lg(\alpha)=\ll\alpha^{1},\alpha^2,\ldots,\alpha^{\lh(\alpha)}\rr$ of length $\lh(\alpha)$ of timestamps in ${T}$ with $\alpha^{i}\neq\alpha^{j}$ for ${i}\neq{j},$ and such that
  for all $\alpha,\beta\in\proc({R})$ and ${i}\leqslant\lh(\alpha)$ the following three axioms are satisfied:
  \begin{itemize}
  \item [1.\,]
    If $\ll n, \alpha \rr\in{T},$
    then ${n}\leqslant\lh(\alpha)$ and $\alpha^{n}=\ll n, \alpha \rr.$
  \item [2.\,]
    If $\ll n, \alpha \rr\in{T},$
    then $\rf(\ll n, \alpha \rr)=\alpha^{j}$ for some ${j}\leqslant{n}.$
  \item [3.\,]
    If $\alpha^{i}=\ll m, \beta \rr,$
    then $\LG_{m}(\beta)\subseteq\LG_{i}(\alpha),$

    where $\LG_{k}(\gamma)$ is the set $\{\gamma^1,\ldots,\gamma^{k}\}$ for $\gamma\in\proc({R})$ and ${k}\leqslant\lh(\gamma).$
  \end{itemize}
\end{defi}

Then, the Causal Tree that RCT produces is the directed graph $\big \ll {T}, \{ \ll {t}, \rf({t})\rr\,{:}\,{t}\,{\in}\,{T}\} \big \rr$.
Note that formally the notation $\alpha^{i}$ is an abbreviation for
$(\lg(\alpha))_{i}$, i.e. the ${i}\nos$-th term in the sequence $\lg(\alpha),$ and this notation can be used only for $\alpha\in\proc({R})$ and ${i}\leqslant\lh(\alpha).$
Note also that $\alpha^{n}=\ll n, \alpha \rr$ holds only if $\ll n, \alpha \rr\in{T}.$
Position $n$ in the log of $\alpha$ might be taken by somebody's else op  $\ll m, \beta \rr$;
then, $\ll n, \alpha \rr \notin {T},$ see Sec.~\ref{RCTExample}.


\begin{nota}
Suppose that ${R}=\ll{T},\vl,\rf,\lg\rr$ is an RCT, ${t}\in{T},$ and  $\alpha\in\proc({R}).$ Then
  \begin{itemize}
  \item [\ding{46}\,]
    $\nos\ndx_\alpha({t})\coloneq$ ${i}$ such that ${t}=\alpha^{i}, $
    if ${t}\in\LG_{\lh(\alpha)}(\alpha);$
  \item [\ding{46}\,]
    $\nos\ndx_\alpha({t})\coloneq\infty,\quad$
    if ${t}\notin\LG_{\lh(\alpha)}(\alpha).$
  \end{itemize}
\end{nota}

We call $\ndx_{\alpha}({t})$ the \emph{index} of the op ${t}$ in the log of process $\alpha.$
It may be different from its author's index $\andx(t)$ due to different op orders in these logs
(see Fig.~\ref{ChronofoldExample} for examples).

\begin{rema}\label{rem.1}
  Suppose that ${R}=\ll{T},\vl,\rf,\lg\rr$ is an RCT, ${t}\in{T},$ $\beta\in\proc({R}),$ ${k}\leqslant\lh(\beta),$ and $\ll{i},\alpha\rr\in{T}.$  
  Then:
  \begin{itemize}
  \item [\ding{43}\,]
    $\nos\andx({t})=\ndx_{\auth({t})}\big({t}\big).$
  \item [\ding{43}\,]
  $\nos\andx(\beta^{k})\leqslant{k}.$%
  \item [\ding{43}\,]
    $\nos\andx({t})\leqslant\ndx_\beta({t})\quad$ for all $\beta\in\proc({R}).$
  \item [\ding{43}\,]
    $\nos\andx({t}) > \andx(\rf({t})),\quad$  if ${t} \ne \rf({t}).$
  \item [\ding{43}\,]
    $\nos\andx(\ll i, \alpha \rr) = i > \andx( \alpha^j )\quad$ for all $j<i.$  \hfill$\qed$
  \end{itemize}
\end{rema}

Note that an op's index in its author's log is the lowest index it has, in any log.
Also, the index of an op in its author's log is greater than author indices of any preceding ops in the same log,
including its CT parent.
Even with subjective ordering, these features hold because of causal consistency and the way we defined timestamps.
We will rely on these features later.

\begin{defi}[\textbf{Causal partial order $\sqsubseteq$}]
  Suppose that ${R}=\ll{T},\vl,\rf,\lg\rr$ is an RCT and ${s},{t}\in{T}.$ Then:

  \begin{enumerate}
  \item [\ding{46}\,]
    $\nos{s}\sqsubseteq{t}\quad$ iff \quad
    there is a sequence $\ll{r}_1,\ldots,{r}_{n}\rr$ such that
    \begin{itemize}
    \item       $\nos{r}_1={s},$
    \item       $\nos{r}_{n}={t},\quad$ and
    \item       $\nos{r}_{i}=\rf({r}_{{i}+1})\ $ or
      $\:{r}_{i}\in\LG_{\andx({r}_{{i}+1})}\big({\auth({r}_{{i}+1}})\big)\quad$
      for all ${i}<{n}.$
    \end{itemize}
  \item [\ding{46}\,]
    $\nos{s}\sqsubset{t}\quad$ iff $\quad{s}\sqsubseteq{t}$ and ${s}\neq{t}.$
  \end{enumerate}
\end{defi}

\begin{rema}[\textbf{Consistency of $\andx$ and causality}]
  Suppose that ${R}=\ll{T},\vl,\rf,\lg\rr$ is an RCT, ${s},{t}\in{T},$ and
  ${s}\sqsubset{t}.$ Then $\andx({s})<\andx({t}).$\hfill$\qed$%
\end{rema}

\begin{lemm} \label{lem.causal}
  Suppose that ${R}$ is an RCT, $\alpha\in\proc({R}),$ ${i}\leqslant\lh(\alpha),$ and $\rf(\alpha^{i})=\ll k, \gamma \rr.$
  Then:
  \begin{itemize}
  \item [\ding{43}\,]
    $\nos\LG_{k}(\gamma)\subseteq\LG_{i}(\alpha).$
  \end{itemize}
  That is, $\nos\LG_{\andx(\rf(\alpha^{i}))}\Big(\auth\big(\rf(\alpha^{i})\big)\Big)
    \subseteq\LG_{i}(\alpha).$\hfill$\qed$%
\end{lemm}

\begin{corr}[\textbf{Causal closedness of $\LG_{i}(\alpha)$}]
  Suppose that ${R}$ is an RCT, $\alpha\in\proc({R}),$ and ${i}\leqslant\lh(\alpha).$
  Then the set $\LG_{i}(\alpha)$ is causally closed.
  That is, if ${s}\sqsubseteq{t}$ and ${t}\in\LG_{i}(\alpha),$
  then ${s}\in\LG_{i}(\alpha).$\hfill$\qed$%
\end{corr}

%
%
%

Historically, CT used at least five different timestamping schemes (Lamport~\cite{lamport1978ordering}, hybrid~\cite{demirbas2014logical}, abbreviated/char tuple~\cite{grishchenko_2010}, calendar based and others).
Given their role in the system, even subtle details had a lot of impact.
The scheme defined here is named \emph{log timestamps}.
Instead of incrementing the value of the greatest timestamp seen, like the Lamport scheme does, we set it to the op's index in its author's log.
The lexicographic ordering of log timestamps is compatible with the causal order (Rem.~\ref{rem.1}).
In addition to that, it also provides a lower bound for the op's index in any log.
Pragmatically speaking, it is the same number most of the time, as the level of contention between replicas of a text tends to be small.
That makes it possible for us to switch from log timestamps to log indices and back, with very little friction.

The importance of this becomes clear when we consider our two uses for timestamps: referencing operations and forming the ATO.
Locally, referencing by index is much more convenient.
The convenience of using an index for ordering depends on whether it matches the author's index.
If $\alpha^n=\ll n, \beta \rr$ then the index also equals the most lexicographically significant part of the timestamp.
If not, it provides an upper bound on that part of the timestamp due to Rem.~\ref{rem.1}.
That is enough to determine the ATO in the absolute majority of the cases.
So most of the time, an RCT implementation may use indices instead of timestamps.

Given that, a process may convert logical timestamps to its log indices once it receives an operation from another process.
Then, it proceeds with the indices.
When sending operations  out, it performs the reverse conversion.
This way, we \emph{insulate} data structures from the complexity of a distributed environment.

Another important feature of log indices is their stability.
The main source of the famous OT complexity is its reliance on a linear addressing system that is not stable between edits.
We avoid that here, to our great advantage, see Sec.~\ref{costruct}.
Given all of the above, we believe it is as natural to use log timestamps for versioned texts as using quaternions for 3D graphics and modeling.
In the next section, we introduce a versioned text data structure that is comparable to plain-text data structures in terms of complexity and overheads.

\section{Chronofold} \label{chronofold}

Every data structure for versioned text has its advantages and shortcomings.
A plain text string is the most simple and the most extensively used data structure in the world.
Unfortunately, a string is edited by splicing;
once we insert a character in the middle, we have to rewrite half the text.
Hence, text edits are  $O(N)$ while comparing (diffing) two versions of a string by the Myers algorithm is $O(ND)\nos$.
Weave is the most natural data structure for diffing and merging, but editing a weave also requires splicing.
A log is an append-only data structure, hence has $O(1)$ edits.
But, recovering a text from a log of edits is not trivial.
Notably, log is a \emph{persistent} data structure in the sense that every prefix of a log is also its complete version.
Similarly, any postfix of a log is a list of recent changes, which is very convenient for replication and synchronization.
Piece tables as used by many text editors have either $O(1)$ or $O(\log N)$ edits and may provide very limited versioning functionality.

\begin{figure}[t]
	\caption{Versioned text: ``MINSK'' corrected to ``PINSK'', stored in different data structures ($\emptyset$: root, $\triangleleft$: tombstone).}
	\centering
	\includegraphics[width=0.32\textwidth]{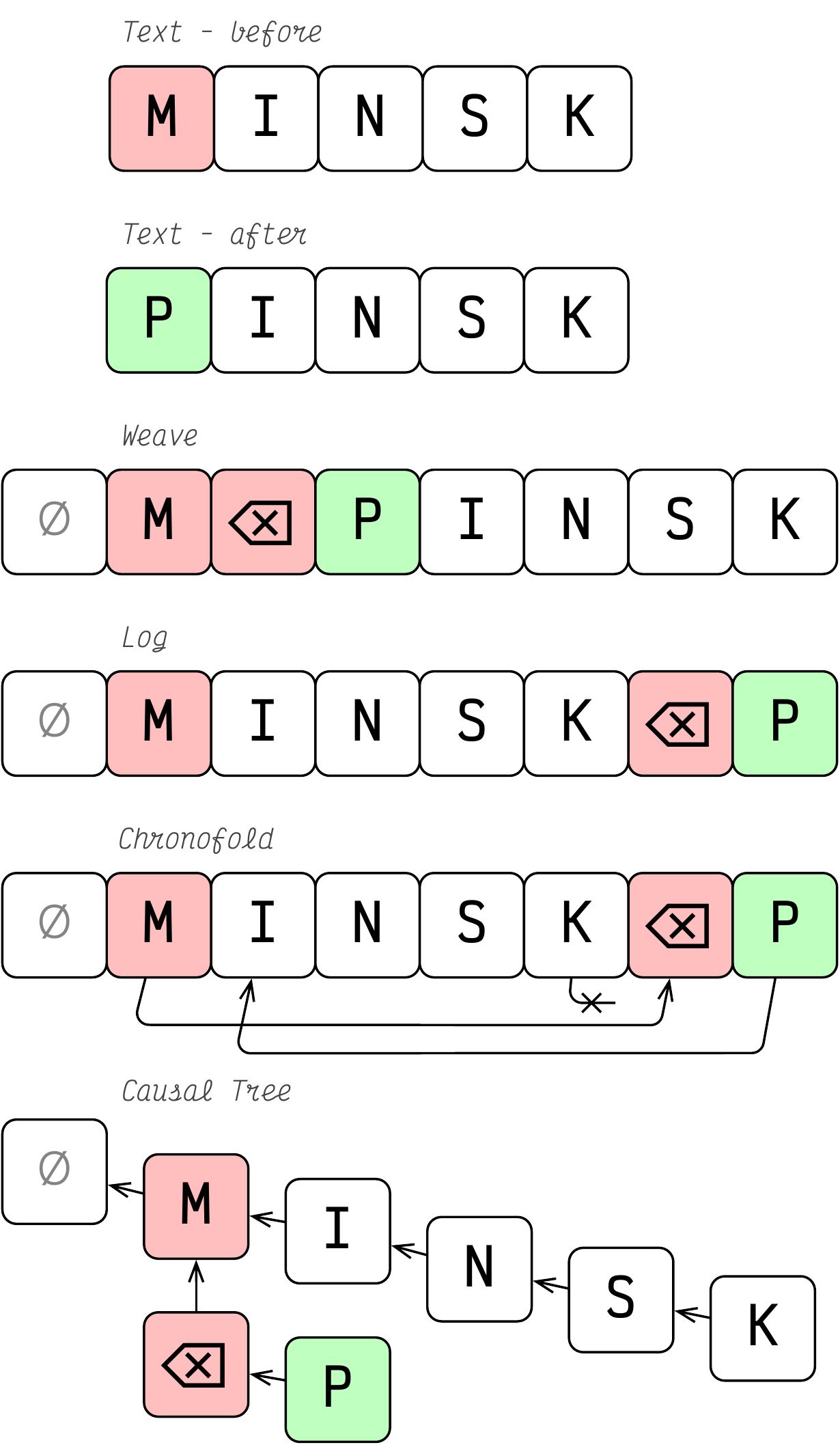} \label{datastr}
\end{figure}


So ideally, we want a replicated versioned text data structure that is stored in an array, addressed by indices, needs no splicing, allows access to past versions of the text and merge of remote changes.
We achieve that by combining a weave and a log:
a \emph{chronofold}\footnote{The name of the data structure was decided by a popular vote~\cite{vote}.}
 is a subjectively ordered log of tuples $\ll \vl(\alpha^i), \ndx_\alpha(\w(\alpha^i)) \rr,$
$i \leqslant \lh(\alpha)$, where $\w(\alpha^i)$ is the operation following $\alpha^i$ in the weave.
So, the second element of the tuple forms a linked list that contains the weave and thus any version of the text.
In the C notation, a text chronofold may look like:

\begin{verbatim}
  struct {
      char32_t codepoint; // UTF-32 character
      uint32_t next_ndx; // weave linked list
  } * cfold;
\end{verbatim}

By reading a chronofold like a log, we see the history of changes.
By reading it along the linked list, we may see any version of the text.
A chronofold has the good properties of a log, a weave and a piece table: it is splicing-free, versioned and very convenient for synchronization.
A chronofold entry takes less space than an op due to the absence of timestamps.
We further optimize that in Sec.~\ref{costruct}.

The algorithm for merging new ops into a chronofold resembles well-known CT/RGA algorithms~\cite{ROH2011354,grishchenko_2010,shapiro2011comprehensive}.
Once process $\alpha$ receives an op $\ll i, \beta \rr$, it appends an entry to its chronofold.
Next, it has to find the op's position in the weave and relink the linked list to include the new op at that position.
It locates the new op's CT parent $\rf(\ll i, \beta \rr) = \ll k, \gamma \rr = \alpha^j$ at the index $j$ in the local log.
Here, $k < i$ and $k \le j;$ most of the time we simply have $j=k$.
It inserts the op after its parent, unless it finds \emph{preemptive CT siblings} at that location
(those are ops with greater timestamps also having $\ll k, \gamma \rr\nos$ as their parent).
If found, the new op is inserted after preemptive siblings and their CT subtrees.

If explained in RGA terms~\cite{ROH2011354}, the CT parent becomes ``left cobject'' while preemptive siblings become ``succeeding nodes'' with greater timestamps/vectors.
In the terms of the 2010 paper~\cite{grishchenko_2010}, those are ``parent'' and ``unaware siblings''.
In plain words, preemptive siblings correspond to concurrent insertions into the same point of the text.
It is the trickiest part of CT/RGA, as it is an algorithm-within-an-algorithm with somewhat different properties~\cite{kleppmann2019interleaving}.
Still, it is considered a worthy tradeoff because it keeps other parts simple.
In real-world usage, preemptive siblings are very rare.

Note that the chronofold building algorithm uses information that is not included into the chronofold itself.
Namely, that is the tree-forming $\rf$ relation and the timestamp-to-index mapping $\ndx_\alpha: \ll k, \gamma \rr \to j$.
It may also need $\ndx_\alpha^{-1}: j \to \ll k, \gamma \rr$ for the case of preemptive siblings.
Exporting edits to other replicas needs $\ndx_\alpha^{-1}$ to produce the timestamps.

Importantly, if we only edit a text locally then a chronofold itself suffices.
Namely, as per Remark~\ref{rem.1}, the timestamp of a new locally authored op is greater than other timestamps in the log.
That excludes the case of preemptive siblings, so $\ndx_\alpha^{-1}$ is not needed.
The index of the preceding character should be already known, so $\ndx_\alpha$ is not needed either.

Then, the data for $\rf, \ndx_\alpha, \ndx_\alpha^{-1}$ can be kept in a separate data structure thus removing it from the hot code path.
From the perspective of a text editor, that makes perfect sense: it merges an op once, then reads many times.
This is exactly the insulation layer we mentioned earlier.

The simplest way to store that metadata is to keep a \emph{secondary} log of op timestamps and their $\rf$ indices.
To implement $\ndx_\alpha^{-1}: j \to \ll k, \gamma \rr$ we simply read that log at the index $j$.
To implement $\ndx_\alpha: \ll k, \gamma \rr \to j$, we may need to scan it from position $k$ to the end, in the worst case.
That costs $O(N)$ and from that perspective we might be tempted to store that mapping in a hash map.
That would solve the problem on paper but, as it was described earlier, that may not be a good idea in practice.
One way to avoid those worst-case scans is to keep a separate sorted table of index \emph{shifts}.
Namely, once $\ndx_\alpha(\ll i, \beta \rr)-i>T$ for some threshold value $T$, make a shift table entry  $\sh_\alpha : \ll i, \beta \rr \to \ndx_\alpha(\ll i, \beta \rr) - i$.
Having that entry, we will know that for $\ll j, \beta \rr$ if $j \geqslant i$ then  $\ndx_\alpha(\ll j, \beta \rr) \geqslant \sh_\alpha(\ll i, \beta \rr) + i$.
As long as this correction keeps us within $T$ steps from the target, we do not need additional entries for $\beta$.
This technique is improved in Sec.~\ref{costruct}.

With a shift table, $\ndx$ has complexity $O(\log N)$, which means $O(\log N)$ insertions, except for one adversarial scenario.
Namely, if one op has $O(N)\nos$ CT childen which are fed into our replica in the reverse timestamp order.
Then, the case of preemptive siblings turns into the bubble sort algorithm: $O(N)\nos$ per op, $O(N^2)\nos$ total.
The scenario corresponds to lots of concurrent insertions into the same point in the text.
Due to the properties of causal consistency, one author can not send ops out-of-order, see Def.~\ref{main.def}, Lemma.~\ref{lem.causal}.
So, this scenario should probably include a Sybil attack~\cite{Sybil} too.
There is another chronofold-building algorithm that lacks this unfortunate corner case;
we have to skip it as it depends on many techniques not explained here.

\begin{figure*}
	\includegraphics[width=\textwidth]{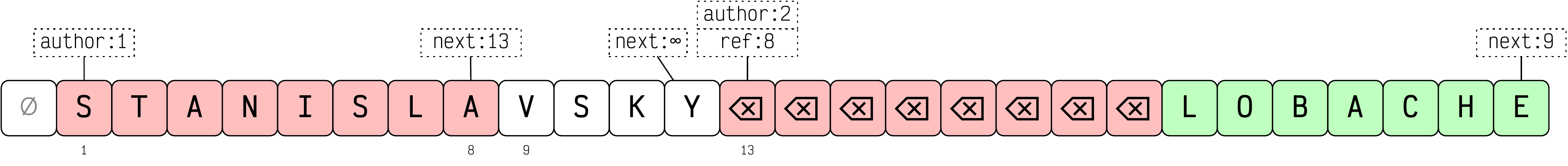}
	\caption{Chronofold and co-structures:
	"STANISLAVSKY" written by author 1 corrected to "LOBACHEVSKY" by author 2.
	Co-structures store timestamps (author, shift), the weave (next), and the tree-forming relation $\rf.$
	Most values are implicit.} \label{lobachevsky}
\end{figure*}

To illustrate what we achieved here, let us consider two typical versioning operations:
recovering a past version and deriving a difference of two versions.
Having a CT weave, we would need timestamps and version vectors to filter non-effective ops (``dead'', ``yet-unborn'') and produce a version of a text or a difference thereof.
Having a chronofold, we may iterate its linked list while ignoring all the ops past certain index.
This way, we produce a version or a diff using indices only.
(Albeit, this only applies to the versions we observed in our subjective linearization of the history; to work with other linearizations we have to build their respective chronofolds.)
As we have mentioned earlier, local editing does not use timestamps either.
That should make CRDT overheads acceptable for the use cases of undo/redo, real-time collaborative editing, or full-scale revision control.

But, whether we speak of editors, collaborative editors or revision control systems,
there is more than plain text.
There is also formatting, highlighting, annotations, versioning.
In this regard, log timestamps make the data structure extremely flexible and adaptable,	see Sec.~\ref{costruct}.

\section{ Co-structures } \label{costruct}

All but the most basic editors overlay the text with various kinds of formatting.
That might be syntax highlighting, spelling errors, compiler warnings, authorship and versioning info, annotations, etc.
Differently from embedded markup (e.g. HTML), overlays are decoupled from the text stream, as they merely \emph{reference} text ranges.
Sometimes, the code responsible for such overlays may be computationally expensive, so it runs asynchronously in separate threads, processes or remotely on servers.
Sometimes, these overlays are stored separately.
Either way, as the text keeps changing, the referenced ranges become slightly off.
Effectively, some editors have to run miniature OT engines to correct for that effect.

Fortunately, log indices create a stable addressing system for the edited text.
As long as we stay with the same replica and same linearization, the indices are not affected by edits.
That lets us build \emph{co-structures}, overlay data structures linked to the text through log indices.
Co-structures reference text ranges, but instead of text positions they use log indices,
so no correction needed.
This is an improvement over past CT editor engines that used logical timestamps to denote such ranges.
Again, we evaded the use of distributed primitives.

As a simple example, we may track a binary attribute by keeping a bitmap (e.g. whether a letter is bold).
For richer attributes we may use a vector, etc.
Although, keeping track of individual characters may not be the most convenient approach.
In case we need to reference character ranges, one possible data structure is a \emph{range map}.
Namely, we divide a chronofold into a number of semi-intervals $[a_i,b_i)$ so $a_i=b_{i-1}, i \ne 0$.
Each interval has uniform formatting $f_i$.
That formatting we keep in a sorted map $a_i \to f_i$.
When iterating the chronofold in the weave order, we check the range map for formatting changes.

As text editors tend to operate in (row,column) coordinates, we may dedicate another co-structure to that purpose.
Namely, a Table of Contents (ToC) listing log indices of all the effective newlines of the text in their weave order.
Having that, we can start iteration from an arbitrary line's beginning.
This way, we may avoid storing the plain text as a separate structure.
Instead, we may produce any line of the text on-demand by scanning the respective piece of the weave.
Again, co-structures make a chronofold extremely flexible.

Note that a slightly out-of-date co-structure can still be applied to the chronofold if that makes sense.
As co-structures are decoupled from the chronofold, they can be (de)activated, (re)stored, (re)built and/or updated, all independently from the editing process.
The only limitation is that the subjective order must stay the same.

Interestingly, this co-structure technique may serve to optimize the chronofold itself.
In part, we already did that in the Sec.~\ref{chronofold} by offloading metadata to a secondary log.
As a next step, we may offload \texttt{next\_ndx} pointers to a co-structure of their own.
A typical verioned text consists of \emph{spans} of sequentially typed characters: words, sentences, deltas.
Simultaneous typing in a real-time collaborative editor may produce messier patterns.
But, based on our experience with deployed systems, that is a rare exception, not the rule.
In a typical chronofold, most of \texttt{next\_ndx} pointers would be equal to $i+1$.
Instead of spending memory for every such value, we may offload them to a separate sparse array, where only non-trivial pointers are mentioned.
In the resulting implementation, a \emph{thinned} chronofold is simply a log of UTF-32 codepoints.
As yet another optimization, we may notice that non-BMP codepoints are very rare.
If so, we may reduce the core chronofold to a UTF-16 string where all non-BMP codepoints are marked with a special value and stored in a yet another sparse array.

Let's return to the secondary log carrying op timestamps and $\rf$ indices (Sec.~\ref{chronofold}).
Author's indices tend to match our local log indices in practice.
Even if not, spans will have the same index \emph{shift} due to concurrent edits present in our log before the span.
So, instead of individual timestamps we may store timestamp ranges in two range maps (authors and shifts respectively), thus avoiding per-character metadata.
In this case, $\ndx_\alpha^{-1}$ takes $O(\log N)$ cycles as it only needs two range map lookups.
$\ndx_\alpha$ is formally $O(N)$ as it may potentially need to scan a range map to find an op that was shifted from beginning to end.
Optimizing this case is possible, but hardly worth it in non-adversarial scenarios,
as we need $ndx_\alpha$ for head-of-span insertions only.

As the final optimization in this paper, we will use the fact that all the co-structures are addressed by the log index.
That means, it is possible to put several of them into the same container, to amortize costs.
Fig.~\ref{lobachevsky} shows a chronofold with its secondary log and weave pointer co-structures.
These can be stored in the same sorted map, assuming the integer key has two flag bits to differentiate between co-structures.

Our span-friendly technique can be even more effective if we discard the history once it becomes irrelevant.
In \texttt{git} terms, that is called a \emph{rebase}, a mandatory procedure in larger projects.
As an extreme case, if a history of a document is discarded entirely, then the text is represented as a single sequential insertion, with no tombstones.
In such a case, a text, a weave and a chronofold become the same sequence.
The \texttt{next\_ndx}, \texttt{author} and \texttt{shift} co-structures will have one entry each.
If we start editing such a rebased text, a chronofold would look very much like a piece table: an initial snapshot and a separate log for the new edits.

This way or another, if the cost of co-structures is sifficiently amortized,
a chronofold's footprint becomes much closer to that of a piece table,
or a plain non-versioned UTF-16 string, as used in Java, JavaScript, etc.
That makes it useful as a general-purpose data structure for versioned text.

\section{ Conclusion } \label{done}

As a data structure, chronofold addresses the shortcomings of weave-based CRDTs.
It is a simple array-based data structure with $O(1)$ inserts that might work faster than a plain string in many cases.
It works like a piece table for editing, like a log for replication, and like a weave for versioning.
The authors are looking forward to see chronofold applications in the domains of revision control systems, collaborative software	and development environments.

The greatest surprise to the authors though is that linear addressing is applicable to a partially ordered system.
The concepts of log timestamps and subjective linear orders mitigate the cognitive and computational costs of a distributed data model.
That may potentially find applications beyond the domain of text versioning.

\section{Acknowledgements}

The work was supported by the Protocol Labs RFP program.
Authors thank Anton Lipin, Ilya Sukhopluev, Maxim Shafirov, and
Yuriy Syrovetskiy for sharing their reflections and feedback.

\bibliographystyle{plain}
\bibliography{chronofold}{}

\appendix

\section{RCT Example} \label{RCTExample}

\newcommand{\alice}{\alpha}
\newcommand{\bob}{\beta}
\newcommand{\carol}{\gamma}

\begin{figure}[h]
	\centering
  \includegraphics[width=0.32\textwidth]{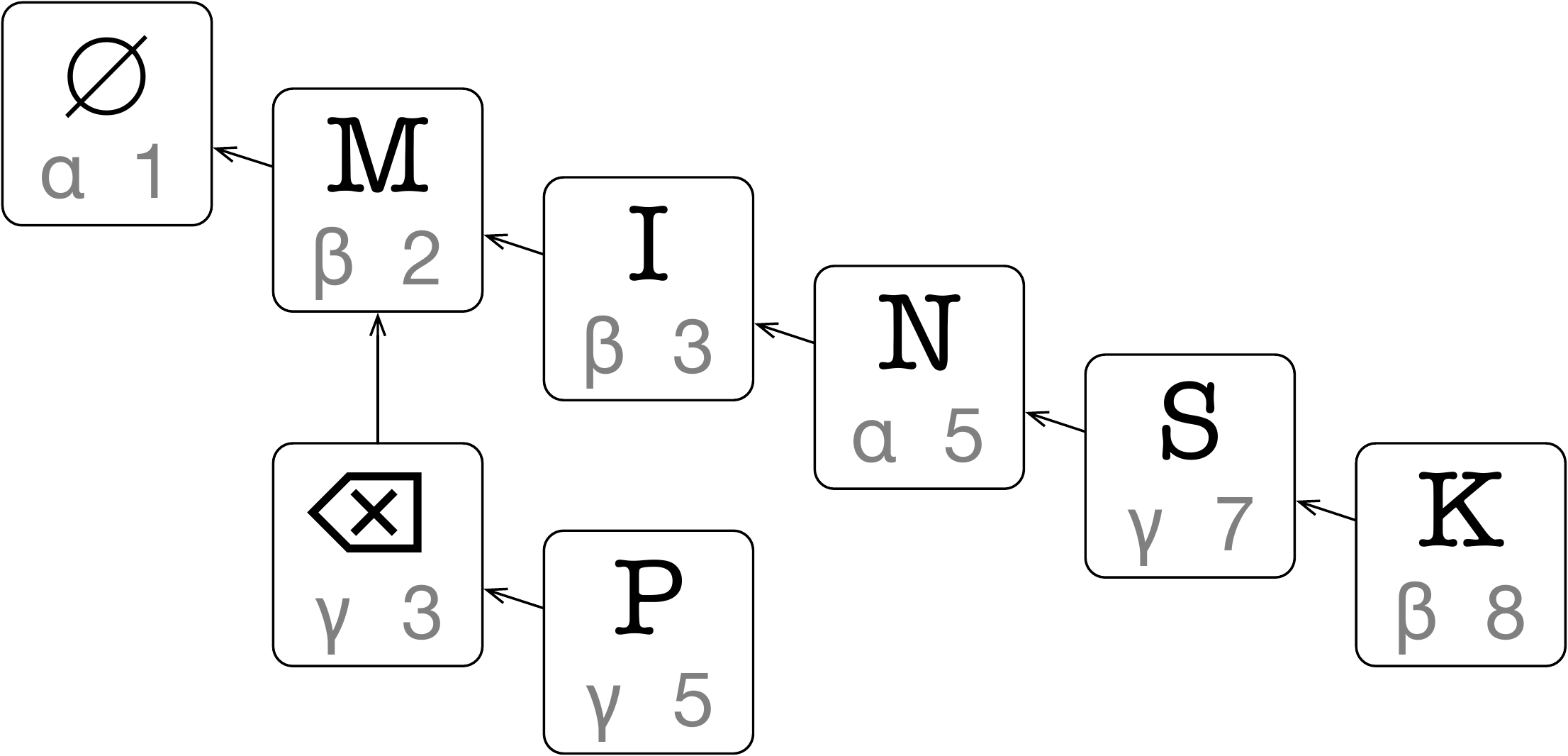} 
  \caption{Three laborous editors struggle to spell ``PINSK''} 
\end{figure}

  In the above RCT example, three processes are named Alice ($\alice\nos$), Bob ($\bob\nos$), and George ($\carol\nos$).
  Alice produced ops
  $$
  \big\ll\ll1,\alice\rr,\ll1,\alice\rr,\varnothing\big\rr
  \text{ and }
  \big\ll\ll5,\alice\rr,\ll3,\bob\rr,N\big\rr\:;
  $$
  Bob produced ops
  $$
  \big\ll \ll2,\bob\rr, \ll1,\alice\rr, M \big\rr,\:
  \big\ll \ll3,\bob\rr, \ll2,\bob\rr, I \big\rr,
  \text{ and }
  \big\ll \ll8,\bob\rr, \ll7,\carol\rr, K \big\rr;
  $$
  and finally George produced ops
  $$
  \big\ll \ll3,\carol\rr, \ll2,\bob\rr, \triangleleft \big\rr,\:
  \big\ll \ll5,\carol\rr, \ll3,\carol\rr, P\big\rr,
  \text{ and }
  \big\ll \ll7,\carol\rr, \ll5,\alice\rr, S\big\rr.
  $$
  
  Then we have the following Replicated Causal Tree ${R}=\ll{T},\vl,\rf,\lg\rr.$
  The set of timestamps ${T}$ is
  $$\big\{
  \ll1,\alice\rr,
  \ll5,\alice\rr,
  \ll2,\bob\rr,
  \ll3,\bob\rr,
  \ll8,\bob\rr,
  \ll3,\carol\rr,
  \ll5,\carol\rr,
  \ll7,\carol\rr
  \big\};
  $$
  the function $\vl$ has the following values on the above timestamps (written in the same order):
  $$
  \varnothing,\:
  N,\:
  M,\:
  I,\:
  K,
  \triangleleft,\:
  P,\:
  S\,;
  $$
  the function $\rf,$ that defines the places of application of the above characters, has the following values:
  $$
  \ll1,\alice\rr,\,
  \ll3,\bob\rr,\,
  \ll1,\alice\rr,\,
  \ll2,\bob\rr,\,
  \ll7,\carol\rr\,
  \ll2,\bob\rr,\,
  \ll3,\carol\rr,\,
  \ll5,\alice\rr\,.
  $$
  The set of processes of our RCT is $\proc({R})=\{\alice,\bob,\carol\}.$
  The logs of these processes depend on the order in which they received ops from each other. For example, the logs could be the following:
  $$
  \lg(\alice)=
  \big\ll
  \alice^1,\alice^2,\alice^3,\alice^4,\alice^5,\alice^6,\alice^7,\alice^8
  \big\rr=
  $$
  $$
  =\big\ll
  \ll1,\alice\rr,
  \ll2,\bob\rr,
  \ll3,\carol\rr,
  \ll3,\bob\rr,
  \ll5,\alice\rr,
  \ll5,\carol\rr,
  \ll7,\carol\rr,
  \ll8,\bob\rr
  \big\rr;
  $$
  $$
  \lg(\bob)=\big\ll
  \ll1,\alice\rr,
  \ll2,\bob\rr,
  \ll3,\bob\rr,
  \ll3,\carol\rr,
  \ll5,\carol\rr,
  \ll5,\alice\rr,
  \ll7,\carol\rr,
  \ll8,\bob\rr
  \big\rr;
  $$
  $$
  \lg(\carol)=\big\ll
  \ll1,\alice\rr,
  \ll2,\bob\rr,
  \ll3,\carol\rr,
  \ll3,\bob\rr,
  \ll5,\carol\rr
  \ll5,\alice\rr,
  \ll7,\carol\rr
  \big\rr\,.
  $$
  
  In this example $\lh(\alice)=\lh(\bob)=8$ (because both processes $\alice$ and $\bob$  have received all eight ops) and $\lh(\carol)=7$ (since the process $\carol$ has not received the op with id $\ll8,\bob\rr$).
  
  The fourth timestamp in the log of process $\alice$ is $\alice^4=\ll3,\bob\rr.$
  The author of this timestamp is $\bob$: we have
  $\,\nos\auth(\alice^4)=$
  $\nos\auth(\ll3,\bob\rr)=\bob.\,$
  The author's index of this timestamp is $3$: we have
  $\andx(\alice^4)=\andx(\ll3,\bob\rr)=3.$
  Also we have $\rf(\alice^4)=\ll2,\bob\rr$
  and $\vl(\alice^4)=I.$
  From this information we can restore the op that this timestamp identifies: the op with id $\alice^4$ is
  $\big\ll \ll3,\bob\rr, \ll2,\bob\rr, I \big\rr.$
  
  Also we have
  $$\LG_{4}(\alice)\coloneq
  \{\alice^1,\alice^2,\alice^3,\alice^4\}=
  $$
  $$
  \big\{\ll1,\alice\rr,
  \ll2,\bob\rr,
  \ll3,\carol\rr,
  \ll3,\bob\rr\big\}
  =
  \big\{\ll3,\bob\rr,\ll2,\bob\rr,
  \ll1,\alice\rr,\ll3,\carol\rr
  \big\}
  $$
  --- the order is irrelevant here because $\LG_{4}(\alice)$ is a \emph{set}, not a sequence (unlike $\lg(\alice)$).
  
  It is not hard to verify that the axioms (1)--(3) are satisfied in the above RCT.
  
    Further on, we have
    $$
    \ndx_\alice(\ll3,\carol\rr)=3
    \quad \text{because}\ \ll3,\carol\rr=\alice^3;
    $$
    $$
    \ndx_\bob(\ll3,\carol\rr)=4
    \quad \text{because}\ \ll3,\carol\rr=\bob^4;
    $$
    $$\ndx_\carol(\ll3,\carol\rr)=3
    \quad \text{because}\ \ll3,\carol\rr=\carol^3.
    $$
    Note that $\andx(\ll3,\carol\rr)=3$
    and $\auth(\ll3,\carol\rr)=\carol.$
    Also we have $\ndx_\carol(\ll8,\bob\rr)=\infty$
    because there is no ${i}$ such that
    $\ll8,\bob\rr=\carol^{i}.$

\section{On abiding the rules}

One may wonder, how a process may ensure it abides the axioms in \ref{main.def}.
Even more interestingly, how it can ensure other abide the rules too?

The axiom (3) is satisfied automatically if each process relays the ops in its subjective order.
The axiom (2) is satisfied if every process only creates op that reference previous ops in the log.
To satisfy the axiom (1) each process must properly assign the author's index to each new op.

On the other end of the connection, how a process may ensure the received op does not violate the axioms?
To enforce (1) a process $\alpha$ must keep track of the log indices $\ndx_\beta$ of the process $\beta$ it synchronizes with. 
This way, it may detect out-of-order indices $\andx(t)<\ndx_\beta(t)$ and improperly assigned indices $\auth(t) = \beta, \andx(t) \ne \ndx_\beta(t)$.
To satisfy (2), it should abort synchronization if a newly received op references an op not in the log.

Enforcing (3) transitively as well as op tampering prevention can be achieved by employing  Merkle structures, which is well out of the scope of this paper.

\begin{figure*}
  ~\\
	\includegraphics[height=0.97\textwidth,angle=270]{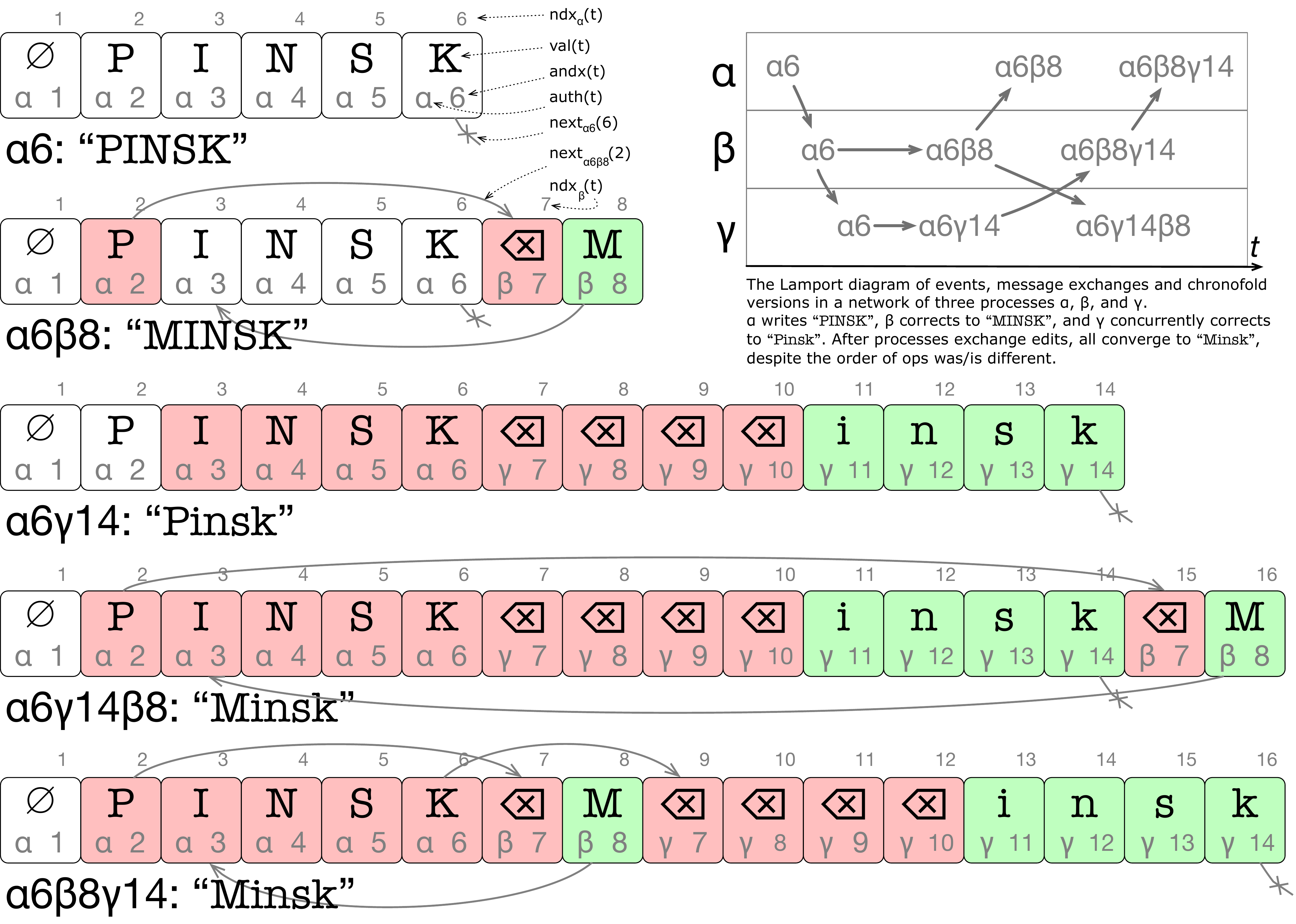}
  \caption{Chronofold Example: three processes, concurrent edits.} \label{ChronofoldExample}
  ~\\
\end{figure*}

\end{document}